\documentclass[aps,prb,preprint,showpacs,superscriptaddress,floatfix]{revtex4}
\usepackage{graphics}
\usepackage{epsfig}
\usepackage{amsmath}
\usepackage{amssymb}
\usepackage{amsthm}
\usepackage{calc}

\newcommand{\beq}{\begin{equation}}
\newcommand{\eeq}{\end{equation}}
\newcommand{\bea}{\begin{eqnarray}}
\newcommand{\eea}{\end{eqnarray}}
\newtheorem*{prop}{Proposition}
\newtheorem*{definition}{Definition}

\begin{document}
\title{Entanglement in fermionic Fock space}
\author{G\'abor S\'arosi and P\'eter L\'evay}
\affiliation{Department of Theoretical Physics, Institute of
Physics, Budapest University of Technology, H-1521 Budapest,
Hungary}
\date{\today}
\begin{abstract}
We propose a generalization of the usual SLOCC and LU classification of entangled  pure state fermionic systems based on the Spin group. Our generalization uses the fact that there is a representation of this 
group
acting
on the fermionic Fock space which
when restricted to fixed particle number subspaces recovers naturally the usual SLOCC transformations. The new ingredient is the occurrence of Bogoliubov transformations of the whole Fock space changing the particle number. The classification scheme built on the Spin group prohibits naturally entanglement between states containing even and odd number of fermions.
In our scheme the problem of classification of entanglement types boils down to the classification of spinors where totally separable states are represented by so called pure spinors. 
We construct the basic invariants of the Spin group and show how some of the known SLOCC invariants are just their special cases.
As an example we present the classification of fermionic systems with a Fock space based on six single particle states. 
An intriguing duality between two different possibilities for embedding three-qubit systems inside the fermionic ones is revealed.
This duality is elucidated via an interesting connection to configurations of wrapped membranes
reinterpreted as qubits.

\end{abstract}
 \maketitle{}

\maketitle

\section{Introduction}

In quantum information theory the concept of entanglement is regarded as a resource for completing various tasks which are otherwise unachieveable or uneffective by means of classical methods\cite{Nielsen}. In order to make use of entanglement in this way one first needs to classify quantum states according to the type of entanglement they possess. A physically well-motivated classification scheme of multipartite quantum systems is based on protocols employing local operations and classical communications (LOCC)\cite{Nielsenletter}. Two states are said to be LOCC equivalent if there is a \textit{reversible} LOCC transformation between them. Obviously reversible LOCC transformations form a group and the aforementioned classification problem boils down to the task of identifying orbits of a group on one of its particular representations. In the case of pure state systems with \textit{distinguishable} constituents 
represented by the Hilbert spaces
$\mathcal{H}_1,\mathcal{H}_2,...,\mathcal{H}_n$ and the composite Hilbert space $\mathcal{H}_1\otimes \mathcal{H}_2\otimes\cdots
\otimes \mathcal{H}_n$ the 
LOCC group is 
simply the group of local unitary transformations  
\beq
U(\mathcal{H}_1)\otimes U(\mathcal{H}_2)\otimes ...\otimes U(\mathcal{H}_n).
\eeq
For practical purposes it is sometimes more convenient to consider the classification of states under reversible stochastic local operations and classical communications (SLOCC)\cite{Bennett,Dur}. For pure state multipartite systems the SLOCC group of such transformations is just the one of invertible local operators (ILO)
\beq
GL(\mathcal{H}_1)\otimes GL(\mathcal{H}_2)\otimes ...\otimes GL(\mathcal{H}_n).
\eeq

For systems with distinguishable constituents on pure state SLOCC classification there is a great variey of results available in the literature. A somewhat more recent question is the one of entanglement classification for pure state systems with \textit{indistinguishable} constituents\cite{Ghir,Sch1,Sch2,Eckert,Vedral,Long,You,levvran1,levvran2,Djokovics,djok1,djok2}. For these ones the Hilbert space is just the symmetric (bosons) or antisymmetric (fermions) tensor power of the single particle Hilbert space $\mathcal{H}$ i.e. for an $n$ particle system $\vee^n \mathcal{H}$ or $\wedge^n \mathcal{H}$ respectively. The indistinguishable SLOCC group is again the linear group $GL(\mathcal{H})$ but this time with the $n$-fold diagonal action on $n$ particle states.

For low dimensional systems with just few constituents the SLOCC classes are well-known. These results have originally been  obtained by mathematicians rediscovered later in the entanglement context by physicists.
For bipartite bosonic and fermionic systems a method similiar to the one based on the usual Schmidt decomposition provides the SLOCC and LU classification\cite{Sch2,Ghir}. For tripartite pure state fermionic systems the SLOCC classification is available up to the case of a 
nine 
dimensional single particle Hilbert space (with the six dimensional case being the first non-trivial one)\cite{levvran1,levsar,levsar2,Reichel,Schouten,Gurevich1,Gurevich2,Vinberg}. 
Some interesting results concerning fourpartite fermionic systems with eight single particle states appeared recently\cite{djok2}.
We note that most of the cases where the complete SLOCC classification is known
correspond to prehomogeneous vector spaces classified by Sato and Kimura\cite{Satokimura,Kimura}.

In this work we introduce a classification scheme valid for fermionic states based on the natural action of the Spin group on the corresponding Fock space. 
Physically the
Fock space of identical particles is the Hilbert space which describes processes where the number of constituents (particles) is not conserved. 
In the case of fermions with a finite dimensional single particle Hilbert space $\mathcal{H}$ the Fock space is just the vector space underlying the exterior algebra over $\mathcal{H}$ and hence is also of finite dimension. It is well known to mathematicians that the Spin group can be represented on this exterior algebra\cite{Chevalley}.
We show that when restricted to fixed particle number subspaces this group action is just the usual one of the indistinguishable SLOCC group. The transformations mixing these subspaces are the so called Bogoliubov 
transformations\cite{Bogol}, well known in condensed matter physics, where these transformations relate different ground states 
of a system usually corresponding to a phase transition. Such an 
example is the usual 
mean-field treatment 
of BCS superconductivity\cite{BCS} where the BCS ground state is related by a Bogoliubov transformation to the Fermi-sea state. 
A characteristic feature of our new classification scheme is that it distinguishes between the even and odd particle states of the Fock space hence such ``fermionic'' and ``bosonic'' states can not be in the same class. 
The novelty in our approach is that unlike in the usual SLOCC scheme now particle number changing protocols are also allowed. 

We emphasize that our proposed classification scheme is precisely the one well-known to mathematicians under the term "classification of spinors". 
In this terminology a spinor is just an element of the vector space on which the two-sheeted cover of the orthogonal group (i.e. the Spin group) is represented.
Two spinors are equivalent if and only if there exist an element of the Spin group which transforms one spinor to the other.
Now the word "classification" means the decomposition of the space of spinors into equivalence classes (orbits) and determining the stabilizer of each orbit, a
problem already solved by mathematicians up to dimension twelve\cite{Igusa}. We will see that from the physical point of view this means that a full classification of fermionic systems according to our proposed scheme is available up to single particle Hilbert spaces of dimension six.
Apart from generalizing the SLOCC scheme one of the aims of the present paper was also to communicate these important results to the physics and quantum information community in an accessible way.

The organization of this paper is as follows. In Section \ref{sec:slocc}. we introduce the action of the Spin group on the fermionic Fock space and describe the above ideas in detail. We then introduce the notion of pure spinors and argue that these are generalized separable states. In Section \ref{sec:locc}. we show that the unitary (probability conserving) subgroup of the complex spin group is its compact real form and that it naturally incorporates the LU group. In Section \ref{sec:momentmap}. we introduce some tools from the theory of spinors such as the Mukai pairing\cite{Mukai}, the moment map\cite{genHitchin} and the basic invariants under the Spin group. We argue that these tools are useful to obtain an orbit classification. In Section \ref{sec:examples}. we discuss some simple examples and show how the above machinery works in practice. The first non-trivial example is the fermionic Fock space with a six dimensional single particle Hilbert space where a full classification is presented according to 
the 
systems 
equivalence\cite{levsar} to a 
particular Freudenthal triple system\cite{Freudenthal,Krutelevich}.
In closing an intriguing duality between two different possibilities for embedding three-qubit systems inside our fermionic ones is revealed.
This duality is elucidated via an interesting connection to configurations of wrapped membranes
reinterpreted as qubits\cite{wrapornottowrap,qubitsfromextra,review}.

\section{Generalization of SLOCC classification}
\label{sec:slocc}

Let $\mathcal{H}$, dim$\mathcal{H}=d$ be the (finite dimensional) one particle Hilbert space. We define the fermionic Fock space of $\mathcal{H}$ by 
\beq
\mathcal{F}=\mathbb{C}\oplus\mathcal{H}\oplus\wedge^2\mathcal{H}\oplus ... \oplus \wedge^d \mathcal{H}.
\eeq
Obiously dim$\mathcal{F}=\sum_{k=0}^{d}\binom{d}{k}=2^d$. Let us introduce the vector space of creation operators $W=\lbrace f^+:\wedge^k\mathcal{H}\rightarrow \wedge^{k+1}\mathcal{H} | f^+ \text{ is linear}\rbrace$ and define the vector space isomorphism between $W$ and $\mathcal{H}$ via the exterior product:
\beq
|v\rangle \wedge |\phi\rangle = (f_v)^+ |\phi\rangle, \; |v\rangle \in \mathcal{H},\;|\phi\rangle \in \mathcal{F}, \;(f_v)^+\in W. 
\eeq
Obviously $(f_v)^+ (f_u)^+=-(f_u)^+ (f_v)^+$
Also the space of annihilation operators $W^*=\lbrace f:\wedge^k\mathcal{H}\rightarrow \wedge^{k-1}\mathcal{H} | f \text{ is linear}\rbrace$ is isomorphic to the dual space, $\mathcal{H}^*$ via the interior product:
\beq
i_{w} |\phi\rangle = f_w |\phi\rangle, \; w \in \mathcal{H}^*,\;|\phi\rangle \in \mathcal{F}, \;f_w\in W^*. 
\eeq
The action of $\mathcal{H}^*$ on $\mathcal{H}$ maps to the canonical anticommutation relations:
\beq
\label{eq:car}
\lbrace f_v,(f_u)^+ \rbrace = v(u) I, \;\; u\in \mathcal{H},\;v\in \mathcal{H}^*,
\eeq
where $I$ denotes the identity operator on $\mathcal{F}$. Indeed, by the antiderivation property of the interior product we have
\beq
i_w(|v\rangle \wedge |\phi\rangle)=(i_w|v\rangle)\wedge |\phi\rangle -|v\rangle \wedge i_v|\phi\rangle =v(w) |\phi\rangle -|v\rangle \wedge i_v|\phi\rangle,
\eeq
which after rearrangement yields \eqref{eq:car}.
Note that if we choose a hermitian inner product $\langle.|.\rangle$ on $\mathcal{H}$ we obtain the \textit{antilinear} inclusion $A: \mathcal{H}\rightarrow \mathcal{H}^*$, $A(v)(u)=\langle v|u\rangle$ which is actually a bijection. In this way the map $f_{A(.)}:\mathcal{H}\rightarrow W^*$ is antilinear and $f_{A(u)}$ is the adjoint of $(f_u)^+$.

Recall that the vector space $V$ is a Clifford algebra Cliff($V$) if there is a multiplication satisfying
\beq
w^2=(w,w)1,\; \forall w\in V
\eeq
with $(\cdot,\cdot)$ an appropriate inner product on $V$.
Now consider the vector space $V=W\oplus W^*$ with the usual operator multiplication and define the inner product with the use of the anticommutator:
\beq
\label{eq:Winnerprod}
\begin{aligned}
(f_1+f_2^+,f_3+f_4^+)I &=\frac{1}{2}\lbrace f_1+f_2^+,f_3+f_4^+\rbrace =\frac{1}{2}\lbrace f_1,f_4^+ \rbrace +\frac{1}{2} \lbrace f_2^+,f_3 \rbrace,\\
f_1,f_3 &\in W^*,\; f_2^+,f_3^+\in W.
\end{aligned}
\eeq
By construction $W\oplus W^*$ with this inner product is a Clifford algebra. We have a natural algebra representation of Cliff($W\oplus W^*$) on $\mathcal{F}$:
\beq
\label{eq:repr}
[f_1\oplus f_2^+]|\phi\rangle=f_1|\phi\rangle + f_2^+ |\phi\rangle, 
\;\; |\phi\rangle\in \mathcal{F},\; f_1\in W^*,\; f_2\in W.
\eeq
Indeed,
\beq
[f_1\oplus f_2^+]^2|\phi\rangle=\lbrace f_1,f_2^+ \rbrace |\phi \rangle = (f_1+f_2^+,f_1+f_2^+)|\phi\rangle.
\eeq

We call \textit{Bogoliubov transformations} the set of automorphisms of $W\oplus W^*$ that keeps the anticommutator, i.e. $O:W\oplus W^*\rightarrow W\oplus W^*$ is a Bogoliubov transformation if
\beq
\label{eq:bogoljdef}
\lbrace O(f_1+f_2^+), O(f_3+f_4^+) \rbrace = \lbrace f_1+f_2^+, f_3+f_4^+ \rbrace.
\eeq
This means that w.r.t our inner product $(O(f_1+f_2^+), O(f_3+f_4^+))=(f_1+f_2^+, f_3+f_4^+)$ so the group of all Bogoliubov transformations is $SO(W\oplus W^*)\cong SO(2d,\mathbb{C})$. The Lie algebra of this group is defined by
\beq
\mathfrak{so}(W\oplus W^*) = \lbrace \mathcal{T}| (\mathcal{T}x,y)+(x,\mathcal{T}y)=0, \;\; x,y\in W\oplus W^*\rbrace.
\eeq
This can be satisfied with the parametrization
\beq
\label{eq:Tdefinieing}
\mathcal{T}
\left(
\begin{array}{c}
(f^i)^+ \\
f_j
\end{array}
\right)
=
\left(
\begin{array}{cc}
A^i_{\;k} & \beta^{il} \\
B_{jk} & -A^l_{\;j}
\end{array}
\right)
\left(
\begin{array}{c}
(f^k)^+ \\
f_l
\end{array}
\right), \;\; B_{ij}=-B_{ji},\;\beta^{ij}=-\beta^{ji},
\eeq
where we have picked the bases $\lbrace e_i \rbrace \subset \mathcal{H}$, $\lbrace e^i \rbrace \subset \mathcal{H}^*$ and denoted the corresponding bases of creation and annihilation operators as $\lbrace f_{e_i}^+\equiv (f^i)^+ \rbrace \subset W$ and $\lbrace f_{e^i}\equiv f_i \rbrace \subset W^*$.
It is well known that we can embed the Lie algebra $\mathfrak{so}(W\oplus W^*)$ into Cliff($W\oplus W^*$):
\beq
x,y\in W\oplus W^*\subset \text{Cliff}(W\oplus W^*)\Rightarrow \frac{1}{4}[x,y]\in \mathfrak{so}(W\oplus W^*).
\eeq 
We can represent $\mathfrak{so}(W\oplus W^*)$ on $\mathcal{F}$ via this embedding and the Clifford algebra representation given in \eqref{eq:repr}. 
In this manner the spinor representation\cite{Chevalley} on $\mathcal{F}$ is defined by
\beq
\label{eq:spinorrep}
[T,\left(
\begin{array}{c}
(f^i)^+ \\
f_j
\end{array}
\right)]=
\left(
\begin{array}{cc}
A^i_{\;k} & \beta^{il} \\
B_{jk} & -A^l_{\;j}
\end{array}
\right)
\left(
\begin{array}{c}
(f^k)^+ \\
f_l
\end{array}
\right).
\eeq
It is not difficult to show that $T$ is implemented by
\beq
\label{eq:T}
T=-B-\beta+ A-\frac{1}{2}\text{Tr}A\cdot I,
\eeq
where
\beq
\begin{aligned}
A\in \text{End}(W), & & A= A^i_{\;j}(f^j)^+ f_i, \\
B\in \wedge^2 W: W\rightarrow W^*, & & B=\frac{1}{2}B_{ij}(f^i)^+ (f^j)^+, \\
\beta \in \wedge^2 W: W^*\rightarrow W, & &\beta=\frac{1}{2}\beta^{ij}f_i f_j,
\end{aligned}
\eeq
This shows that $\mathfrak{so}(W\oplus W^*)=\text{End}(W)\oplus \wedge^2 W^*\oplus \wedge^2 W$. 

The whole spin group can be obtained from the Clifford algebra in the following way
\beq
\label{eq:spingroup}
Spin(W\oplus W^*)=\lbrace O=x_1...x_r| x_i=f_{u_i}+f_{v_i}^+\in W\oplus W^*,\;\;(x_i,x_i)=\pm 1,\;\;r\text{ is even}\rbrace.
\eeq
The finite version of \eqref{eq:spinorrep} is the usual double cover
\beq
OxO^{-1}=\mathcal{O} (x), \;\; x\in W\oplus W^*, \;\; O\in Spin(W\oplus W^*), \;\; \mathcal{O}\in SO(W\oplus W^*),
\eeq
where \eqref{eq:spinorrep} can be recovered by putting $O=e^T$ and $ \mathcal{O}=e^\mathcal{T}$. We note that the exponential map gives only the identity component $Spin_0(W\oplus W^*)$. 
For clarity, we always denote matrices in the vector representation acting on $W\oplus W^*$ with caligraphic letters (e.g. $\mathcal{T},\mathcal{O}$) and operators in the Fock space spinor representation with roman letters (e.g. $T$, $O$).

It is known that the representation constructed in this way is reducible: $\mathcal{F}=\mathcal{F}^+\oplus \mathcal{F}^-$, where $\mathcal{F}^\pm$ are the $\pm 1$ eigenspaces of the volume element  $\gamma_V = 2^{d}\prod_{i=1}^d((f^i)^+f_i-\frac{1}{2})$ of the Clifford algebra (this is just the generalization of the usual $\gamma_5$ matrix). These are simply the even and odd particle subspaces of the Fock space:
\beq
\label{eq:evoddFock}
\mathcal{F}^+=\wedge^{ev}\mathcal{H}\otimes (\wedge^d \mathcal{H}{^*})^{1/2}, \;\;\; \mathcal{F}^-=\wedge^{odd}\mathcal{H}\otimes (\wedge^d \mathcal{H}{^*})^{1/2}.
\eeq
From our perspective this reducibility corresponds to the superselection between ``fermionic'' and ``bosonic'' states. The appearance of the factors of $(\wedge^d \mathcal{H}{^*})^{1/2}$ is explained before Eq. \eqref{eq:slocc}. 

Let us now consider the Fock vacuum $|0\rangle$. By the above it is easy to see that a finite $O\in Spin(W\oplus W^*)$ tranformation acts on this as
\beq
O|0\rangle=e^T|0\rangle = (\det e^A)^{-1/2}e^{-B} |0\rangle = (\det e^A)^{-1/2}(1+B+\frac{1}{2}B^2+...+\frac{1}{\lfloor d/2 \rfloor} B^{\lfloor d/2 \rfloor})|0\rangle,
\eeq
since $A$ and $\beta$ annihilates $|0\rangle$. Notice that if $|0\rangle$ is a Fock vacuum of $f$ then $e^{-B}|0\rangle$ is a Fock vacuum of $OfO^{-1}$. As a consequence the orbit $\lbrace O|0\rangle | O\in Spin(W\oplus W^*)\rbrace$ contains possible Bogoliubov Fock vacuums in $\mathcal{F}$. Moreover for an arbitary state obtained with a definite number of creation operators acting on the Fock vacuum, we have
\beq
\label{eq:purestates}
O(f_1^+ ... f_n^+ |0\rangle)=Of_1^+ O^{-1} O f_2 O^{-1} ... Of_n O^{-1} O|0\rangle = (\det e^A)^{-1/2} \mathcal{O}(f_1^+) ... \mathcal{O}(f_n^+) e^{-B} |0\rangle,
\eeq
thus any selected orbit has the same number of appropiate oscillators excited above the appropiate Bogoliubov vacuum state. 

Notice that the transformations with $B_{ij}=\beta^{ij}=0$ has the form of an ordinary SLOCC transformation on the creation operators. Indeed in this case $O(f^i)^+O^{-1}=G^i_{\;k}(f^k)^+$ where $G\in GL(W)$ is the exponential of the matrix $A^i_{\;j}$. However, because of the $\frac{1}{2}\text{Tr}A$ factor in Eq. (\ref{eq:T}) a fix particle number state picks up a determinant factor:
\beq
\label{eq:slocc}
(f^{i_1})^+...(f^{i_m})^+|0\rangle \mapsto (\det G)^{-1/2}G^{i_1}_{\;k_1}(f^{k_1})^+...G^{i_m}_{\;k_m}(f^{k_m})^+|0\rangle.
\eeq
This justifies the factors of $(\wedge^d \mathcal{H}^*)^{1/2}$ in \eqref{eq:evoddFock}.
Now let us introduce the analogues of separable states.

\begin{definition}
A spinor $|\phi\rangle\in \mathcal{F}$ is said to be pure if the annihilator subspace
\beq
E_\phi=\lbrace x\in W\oplus W^* | x|\phi\rangle =0 \rbrace
\eeq
is a maximal isotropic subspace (i.e. has dimension equal to dim $W$).
\end{definition}

Indeed, $E_\phi$ has to be an isotropic subspace since for all $x\in E_\phi$ we have $0=x^2|\phi\rangle=(x,x)|\phi\rangle$ thus $x$ is null. Clearly the Fock vacuum is a pure spinor since its isotropic subspace consists of all the annihilation operators, $E_0=\lbrace f\in W^*\subset W\oplus W^*\rbrace$. Since if an $x\in W\oplus W^*$ annihilates $|\phi\rangle$ we have $x|\phi\rangle=0 \Rightarrow OxO^{-1}O|\phi\rangle=0$, so the annihilator subspace transforms as $E_{O\phi}=\mathcal{O}E_\phi$. As a consequence states on the same $Spin(W\oplus W^*)$ orbit has the same dimensional annihilator subspace thus every state on the vacuum orbit $\lbrace O|0\rangle | O\in Spin(W\oplus W^*)\rbrace$ is a pure spinor. Moreover all the separable states are pure spinors. Indeed, a state of the form $|\phi\rangle=f_1^+...f_n^+|0\rangle$ is annihilated by
\beq
E_\phi=\lbrace f_1^+,...,f_n^+,f_{n+1},...,f_{\dim W} \rbrace.
\eeq
Notice that a ``bosonic'' Fermi sea state is in the same class as the Fock vacuum. More generally\cite{Chevalley,Gualtieri} any pure spinor can be expressed in the form
\beq
\label{eq:purespinor}
|\phi_0\rangle = \lambda e^B f_{v_1}^+...f_{v_k}^+|0\rangle,
\eeq
where $\lambda\neq 0$ is an arbitary complex number, $B=\frac{1}{2}B_{ij}(f^i)^+ (f^j)^+\in \wedge^2W$ and $v_1,...,v_k$, $k\leq d$ are some lineary independent vectors in $\mathcal{H}$. The corresponding annihilator subspace is just 
\beq
E_{\phi_0}=\lbrace e^Bf_{v_1}^+e^{-B},...,e^Bf_{v_k}^+e^{-B},e^Bf_{u_1}e^{-B},...,e^Bf_{u_{d-k}}e^{-B}| u_i(v_j)=0,\;1\leq i\leq d-k,\;1\leq j\leq k\rbrace,
\eeq
where $u_1,...,u_{d-k}$ are lineary independent elements of $\mathcal{H}^*$.
Because of these properties we propose pure spinors to be the analouges of separable states of ordinary SLOCC classification in the fermionic Fock space. 


\section{Generalization of LOCC classification}
\label{sec:locc}

Notice that if we have a hermitian inner product $\langle .|.\rangle$ on $\mathcal{H}$ then consistency with this requies that a Bogoliubov transformed annihilation operator $OfO^{-1}$ must stay the adjoint of the Bogoliubov transformed creation operator $Of^+O^{-1}$ thus if we have $f^+=f^\dagger$ then we must have
\beq
(OfO^{-1})^\dagger = O(f^\dagger)O^{-1}.
\eeq
This constraint means that the spinor representation $O=e^{T}=e^{-B-\beta+ A-\frac{1}{2}\text{Tr}A\cdot I}$ must be unitary w.r.t. this inner product. One can check that this means that the matrix $A^i_{\;j}$ must be antihermitian and the matrices $B_{ij}$, $\beta^{ij}$ must satisfy $(B^\dagger)_{ij}=-\beta^{ij}$, thus the group of admissable Bogoliubov transformations is restricted to $SO(W\oplus W^*)\bigcap SU(W\oplus W^*)$. 
Notice that though the matrix
\beq
\mathcal{T}=
\left(
\begin{array}{cc}
A^i_{\;k} & \beta^{il} \\
B_{jk} & -A^l_{\;j}
\end{array}
\right)
\eeq
is antihermitian, it does not have to be real so the Bogoliubov transformation can still have complex coefficients. Indeed, define the block matrix
\beq
g=
\left(
\begin{array}{c|c}
0 & I \\ \hline
I & 0
\end{array}
\right).
\eeq
With this the $SO$ property of $\mathcal{O}=e^{\mathcal{T}}$ reads as $\mathcal{O}g\mathcal{O}^T=g$ (see \eqref{eq:bogoljdef}). On the other hand $\mathcal{T}$ is antihermitian so $\mathcal{O}^\dagger=\mathcal{O}^{-1}$. Combine the two to get the following reality condition
\beq
\mathcal{O}^*=g\mathcal{O}g.
\eeq
Now we shall prove that the subgroup of $SO(W\oplus W^*)=SO(2d, \mathbb{C})$ satisfying the above reality condition is just the compact real form $SO(2d, \mathbb{R})$. Define the $2d\times 2d$ unitary matrix
\beq
N=\frac{1}{\sqrt{2}}
\left(
\begin{array}{c|c}
 I & iI \\\hline
 I & -iI
\end{array}
\right).
\eeq
It is easy to check that $N$ diagonalizes $g$:
\beq
\label{eq:Ngg0}
gN=Ng_0,
\eeq
where we have defined
\beq
g_0=
\left(
\begin{array}{c|c}
 I & 0 \\\hline
 0 & -I
\end{array}
\right).
\eeq
\begin{prop}
The map $\mathcal{O} \mapsto S=N^\dagger \mathcal{O} N$ is a group isomorphism from the subgroup $\lbrace \mathcal{O}\in SO(2d,\mathbb{C})|\mathcal{O}^*=g\mathcal{O}g\rbrace$ to the group $SO(2d,\mathbb{R})$.
\begin{proof}
It is obvious that the map giving rise to $S$ is a homomorphism and since $N$ is invertible it is also an isomorphism. Now it is very easy to directly check that $gN^*=N$ and hence $N^T=N^\dagger g$. We have
\beq
S S^T=N^\dagger \mathcal{O} NN^T \mathcal{O}^T N^*=N^\dagger g N^*=I,
\eeq
hence for every unitary $\mathcal{O}$ satisfying $\mathcal{O}g\mathcal{O}^T=g$, $S$ is an element of $SO(2d ,\mathbb{R})$. For the converse, we have to check whether $\mathcal{O}=NSN^\dagger$ is a unitary matrix with the condition $\mathcal{O}g\mathcal{O}^T=g$ satisfied for all $S\in SO(2d,\mathbb{R})$. It is obvious that $\mathcal{O}$ is unitary. For the other write
\beq
\mathcal{O}g\mathcal{O}^T=NSN^\dagger g N^* S^T N^T,
\eeq
but again $gN^*=N$ so
\beq
\mathcal{O}g\mathcal{O}^T=NSN^\dagger NS^T N^T=NS S^T N^T= NN^T=g.
\eeq
\end{proof}
\end{prop}

It is very important that the previously SLOCC type transformations with $B=\beta=0$ are restricted in this case to simple LOCC $U(d)$ because $A$ has to be antihermitian. Actually this seems quite intuitive. Recall that the extra constaints arose from requiring $O=e^T$ to be unitary w.r.t. a fixed inner product. But we know that the role of the inner product of our Hilbert space is to associate probabilities to states. Henceforth the restricted Bogoliubov transformations corresponding to $SO(2d,\mathbb{R})$ preserve probabilities while the whole $SO(2d,\mathbb{C})$ does not. This is exactly the physical difference between LOCC and SLOCC transformations.

\section{The moment map}
\label{sec:momentmap}

There is a canonical antiautomorphism of the exterior algebra $\wedge^\bullet \mathcal{H}$, called the transpose which is the linear extension of the map:
\beq
t: (f_1)^+...(f_k)^+|0\rangle \mapsto (f_k)^+...(f_1)^+|0\rangle=(-1)^{\frac{k(k-1)}{2}}(f_1)^+...(f_k)^+|0\rangle.
\eeq
Define a bilinear product on $\mathcal{F}$ as
\beq
\label{eq:pairing}
(\phi,\psi)=(|\phi\rangle^t\wedge |\psi\rangle)_{\text{top}}\in\mathbb{C},
\eeq
where the subscript top means that one has to take the coefficient of the top component i.e. the number multiplying $(f^1)^+ (f^2)^+ ... (f^d)^+|0\rangle$. In the followings the transformation properties of this product under $Spin(2d,\mathbb{C})$ is of central importance to us so we shall examine it in detail here\cite{genHitchin,Chevalley}.

\begin{prop}
Let $x\in W\oplus W^*$, $|\phi\rangle,|\psi\rangle\in \mathcal{F}$. We have
\beq
(x\phi,x\psi)=(x,x)(\phi,\psi).
\eeq
where the inner product $(x,x)$ is the one defined in \eqref{eq:Winnerprod}.
\begin{proof}
Fix an element $\Omega = f_1 f_2 ... f_d$ of the one dimensional space $\wedge^d W^*$. This is unique up to a scale. It is not difficult to check that
\beq
\Omega (f^1)^+ (f^2)^+ ... (f^d)^+|0\rangle = (-1)^{\frac{d(d-1)}{2}}|0\rangle.
\eeq
Taking the coefficient of the top form is equivalent to writing
\beq
(|\phi\rangle^t\wedge |\psi\rangle)_{\text{top}}|0\rangle=(-1)^{\frac{d(d-1)}{2}} \Omega( |\phi\rangle^t\wedge |\psi\rangle).
\eeq
Now any $\phi$ element of $\mathcal{F}$ can be written as a Clifford algebra element (such as $\phi^{(0)}  I+ \phi^{(1)}_{i}(f^i)^+ + \phi^{(2)}_{ij}(f^i)^+(f^j)^+ + ...$) acting on the vacuum $|0\rangle$. So choose $\Phi,\Psi\in Cliff(W\oplus W^*)$ so that $|\phi\rangle = \Phi |0\rangle$ and $|\psi\rangle = \Psi |0\rangle$. This way we have
\beq
(\phi, \psi)|0\rangle=(-1)^{\frac{d(d-1)}{2}}\Omega \Phi^t \Psi|0\rangle.
\eeq
Now since $x\in W\oplus W^*$ by definition we have $(x\Phi)^t=\Phi^tx$ so
\beq
\begin{aligned}
(x\phi, x\psi)|0\rangle &=(-1)^{\frac{d(d-1)}{2}}\Omega (x\Phi)^t x\Psi|0\rangle \\&= (-1)^{\frac{d(d-1)}{2}}\Omega \Phi^t x^2\Psi|0\rangle \\ &=(x,x)(-1)^{\frac{d(d-1)}{2}}\Omega \Phi^t \Psi|0\rangle \\
&=(x,x)(\phi, \psi)|0\rangle.
\end{aligned}
\eeq
where we used the defining relation of the Clifford algebra for $x^2$.
\end{proof}
\end{prop}
 Using this result and \eqref{eq:spingroup} it is straightforward to see that for $O\in Spin(W\oplus W^*)=Spin(2d,\mathbb{C})$ we have
\beq
(O\phi,O\psi)=\pm(\phi,\psi).
\eeq
If $d=$dim $\mathcal{H}$ is even this is the so called Mukai pairing on the irreducible subspaces $\mathcal{F}^{\pm}$ which is symmetric if $d=4k$ and anti-symmetric if $d=4k+2$. Let $|\phi_n\rangle$ be the part of $|\phi\rangle\in \mathcal{F}$ that lies in $\wedge^n \mathcal{H}$. Then the Mukai pairing explicitly reads as
\beq
(\phi,\psi)=\sum_m (-1)^m |\phi_{2m}\rangle \wedge |\psi_{d-2m}\rangle \in  \wedge^d \mathcal{H}\otimes\wedge^d \mathcal{H}^* =\mathbb{C},
\eeq
if $|\phi\rangle,|\psi\rangle \in \mathcal{F}^+$ and
\beq
(\phi,\psi)=\sum_m (-1)^m |\phi_{2m+1}\rangle \wedge |\psi_{d-2m-1}\rangle \in  \wedge^d \mathcal{H}\otimes\wedge^d \mathcal{H}^* =\mathbb{C},
\eeq
if $|\phi\rangle,|\psi\rangle \in \mathcal{F}^-$. As already mentioned, this satisfies $(O\phi,O\psi)=\pm(\phi,\psi)$ for any $O\in Spin(2d,\mathbb{C})$. When $d=4k+2$ with the use of this invariant bilinear product one can associate elements of the Lie algebra $\mathfrak{so}(2d,\mathbb{C})$ to the elements of the Fock space which can be used for classification of orbits and construction of invariants. So define the moment map 
\beq
\begin{aligned}
\mu & :\mathcal{F} \rightarrow  \mathfrak{so}(2d,\mathbb{C}), \\
    & |\phi\rangle \mapsto \mathcal{T}_\phi,
\end{aligned}
\eeq
as
\beq
\label{eq:moment}
g(\mathcal{T}_\phi,\mathcal{T})=\frac{1}{2}(T\phi,\phi), \;\;\forall \mathcal{T}\in \mathfrak{so}(2d,\mathbb{C}),
\eeq
where $g$ is the Killing form. On $\mathfrak{so}(2d,\mathbb{C})$ we have $g(\mathcal{T}_\phi,\mathcal{T})=2(d-1)\text{Tr}(\mathcal{T}_\phi \mathcal{T})$. Note that invariance of the bilinear product requires $(T\phi,\phi)+(\phi,T\phi)=0$, thus in the case of $d=4k$ where this product is symmetric, the moment map vanishes identically. However, if $d=4k$ the product $(\phi,\phi)$ itself is nonvanishing and it is a good quadratic invariant. Nevertheless when $d=4k +2$ the moment map is a useful tool because $\mathcal{T}_\phi$ has good transformation properties. To see this consider $\mathcal{T}_{O\phi}$, the element associated to $O|\phi\rangle$. By definition for every $\mathcal{T}$ we have
\beq
\begin{aligned}
2(d-1)\text{Tr}(\mathcal{T}_{O\phi} \mathcal{T})&=\frac{1}{2}(TO\phi,O\phi)\\&=
\frac{1}{2}(OO^{-1}TO\phi,O\phi)\\&
=\frac{1}{2}(O^{-1}TO\phi,\phi)\\&
=2(d-1)\text{Tr}(\mathcal{T}_{\phi} \mathcal{O}\mathcal{T}\mathcal{O}^{-1})\\&=2(d-1)\text{Tr}(\mathcal{O}^{-1}\mathcal{T}_{\phi} \mathcal{O}\mathcal{T}),
\end{aligned}
\eeq
thus $\mathcal{T}_{O\phi}=\mathcal{O}^{-1}\mathcal{T}_{\phi} \mathcal{O}$. As a consequence the rank of $\mathcal{T}_{\phi}$ is invariant under the action of $Spin(2d,\mathbb{C})$. Moreover the quantities
\beq
q_k(\phi)=\frac{8^2(d-1)^2}{2}\text{Tr}(\mathcal{T}_{\phi}^k), \;\;k\in \mathbb{N}
\eeq
are invariant homogeneous degree $2k$ polinomials in the coefficients of $|\phi\rangle$.

It is instructive to work out the explicit form of $\mathcal{T}_\phi$. For this, put in the definition \eqref{eq:moment} $\mathcal{T}$ and $T$ from \eqref{eq:Tdefinieing} and \eqref{eq:T} respectively and use the parametrization
\beq
\mathcal{T}_\phi= \frac{1}{8(d-1)}
\left(
\begin{array}{cc}
[A_\phi]^i_{\;k} & [\beta_\phi]^{il} \\
{[B_\phi]}_{jk} & -[A_\phi]^l_{\;j}
\end{array}
\right).
\eeq
After matching coefficients of $A^i_{\;k}$, $B_{jk}$ and $\beta^{il}$ one gets
\beq
\begin{aligned}
 {[A_\phi]}^i_{\;k} &=((f^i)^+ f_k\phi,\phi), \\
 [B_\phi]_{jk} &=(f_j f_k \phi,\phi), \\
 [\beta_\phi]^{il} &= ((f^i)^+(f^l)^+ \phi,\phi).
\end{aligned}
\eeq
Now write $|\phi\rangle\in \mathcal{F}^+$ as
\beq
|\phi\rangle = \sum_{m}\frac{1}{(2m)!} \phi^{(2m)}_{i_1...i_{2m}}(f^{i_1})^+...(f^{i_{2m}})^+|0\rangle.
\eeq
Also define the dual amplitudes $\tilde \phi_{(2m)}$ through
\beq
\phi^{(d-2m)}_{i_1...i_{d-2m}}=\frac{1}{(2m)!}{\tilde \phi_{(2m)}}^{j_1...j_{2m}}\epsilon_{j_1...j_{2m}i_1...i_{d-2m}}.
\eeq
With these notations we have
\beq
\label{eq:momentexpl}
\begin{aligned}
{[A_\phi]}^i_{\;k} &= \sum_m \frac{(-1)^m}{(2m-1)!}\phi^{(2m)}_{k j_2...j_{2m}} {\tilde \phi_{(2m)}}^{i j_2...j_{2m}},\\
 [B_\phi]_{jk} &=\sum_m \frac{(-1)^m}{(2m)!}\phi^{(2m+2)}_{k j i_3...i_{2m+2}} {\tilde \phi_{(2m)}}^{i_3...i_{2m+2}}, \\
 [\beta_\phi]^{il} &= \sum_m \frac{(-1)^m}{(2m-2)!}\phi^{(2m-2)}_{j_1...j_{2m-2}} {\tilde \phi_{(2m)}}^{i l j_1...j_{2m-2}}.
\end{aligned}
\eeq
Formulas for $\mathcal{F}^-$ can be obtained easily by replacing every $2m$ with $2m+1$ while leaving the $(-1)^m$ factors unchanged.

\section{Examples}
\label{sec:examples}

\subsection{Two state system}

It is instructive to work out the trivial example where $\mathcal{H}=\mathbb{C}^2$. The full Fock space is $\mathbb{C}\oplus \mathbb{C}^2 \oplus \wedge^2\mathbb{C}$ with even and odd components:
\beq
\begin{aligned}
\mathcal{F}^+ &=\mathbb{C}\oplus \wedge^2\mathbb{C},\\
\mathcal{F}^- &= \mathbb{C}^2 ,
\end{aligned}
\eeq
which are simply one qubit Hilbert spaces. Take
\beq
\begin{aligned}
|\phi\rangle &=\phi_0|0\rangle + \phi_t (f^1)^+(f^2)^+|0\rangle \in \mathcal{F}^+,\\
|\psi\rangle &=\psi_i (f^i)^+|0\rangle \in \mathcal{F}^-.\\
\end{aligned}
\eeq
An easy calculation shows that the moment maps for these states are
\beq
\begin{aligned}
 \mathcal{T}_\phi= \frac{1}{8}
\left(
\begin{array}{cccc}
-\phi_0 \phi_t & 0 & 0 & -\phi_0^2 \\
0 & -\phi_0 \phi_t & \phi_0^2 & 0 \\
0 & -\phi_t^2 & \phi_0 \phi_t & 0\\
\phi_t^2 & 0 & 0 & \phi_0 \phi_t
\end{array}
\right), & \;\;
 \mathcal{T}_\psi= \frac{1}{8}
\left(
\begin{array}{cccc}
\psi_1 \psi_2 & \psi_2^2 &0  &0  \\
-\psi_1^2 & -\psi_1 \psi_2 &  0&0  \\
 0 &0  & -\psi_1 \psi_2 & \psi_2^2\\
  0&0  & \psi_1^2 & \psi_1 \psi_2
\end{array}
\right).
\end{aligned}
\eeq
Of course both of these square to zero, giving $q_k=0$ for all $k$. Moreover both matrices have either rank two or zero corresponding to the fact that we only have two group orbits: the trivial one with the zero vector and the rest. In the case of $|\psi\rangle$ we have $\beta|\psi\rangle=B|\psi\rangle =0$ thus only the $A$ generators act non-trivially on $\mathcal{F}^-$. This means that we only have to consider the action of $GL(2,\mathbb{C})\subset Spin(4,\mathbb{C})$ which is just the SLOCC group of the trivial one-qubit system. 

Note that both $|\phi\rangle$ and $|\psi\rangle$ are pure spinors. Indeed, we see that $|\psi\rangle$ is manifestly of the form \eqref{eq:purespinor}, while $|\phi\rangle$ can be written as
\beq
|\phi\rangle = \phi_0 \text{exp}\left( \frac{\phi_t}{\phi_0} (f^1)^+(f^2)^+ \right)|0\rangle.
\eeq
We mention that the three state system based on $\mathcal{H}=\mathbb{C}^3$ is also one Spin orbit\cite{Igusa} thus contains only pure spinor states. 

\subsection{Four state system}

Here $\mathcal{H}=\mathbb{C}^4$ and because the dimension is divisible by four we do not have a moment map. In the case of $\mathcal{F}^+$ we parametrize a state as
\beq
|\phi\rangle = \eta|0\rangle + \frac{1}{2} \xi_{ij} (f^i)^+(f^j)^+|0\rangle + \frac{1}{4!} \rho \epsilon_{ijkl}(f^i)^+(f^j)^+(f^k)^+(f^l)^+|0\rangle.
\eeq
We have a quadratic invariant
\beq
(\phi,\phi)=2\eta \rho -2\text{Pf}(\xi),
\eeq
where the Pfaffian of the antisymmetrix matrix $\xi$ is $\text{Pf}(\xi)=\frac{1}{2^2 2!}\epsilon^{ijkl}\xi_{ij}\xi_{kl}$. Pure spinors are the ones with\cite{genHitchin,Igusa} $(\phi,\phi)=0$. In particular for two fermion states $\eta=\rho=0$ the relation $(\phi,\phi)=-2\text{Pf}(\xi)=0$ gives the Pl\"ucker relations which are neccesary and sufficient conditions for a fixed fermion number state to be separable\cite{Kasman}. 
Notice that for normalized states the quantity
$0\leq 64\vert{\rm Pf}(\xi)\vert\leq 1 $ is just the canonical entanglement measure used
for two fermions with four single particle states\cite{Sch2,Eckert}.
Notice that an arbitrary two-qubit state
\beq
\vert x\rangle=\sum_{i,j\in\{0,1\}}x_{ij}\vert i\rangle\otimes\vert j\rangle
\eeq
\noindent
can be embedded into this fermionic system as\cite{levvran2,Djokovics}
\beq
\vert\xi_x\rangle =\sum_{i,j\in\{0,1\}}\xi_{ij}(f^{i+1})^+(f^{j+3})^+\vert 0\rangle.
\eeq
\noindent
Under this embedding the entanglement measure $64\vert{\rm Pf}(\xi)\vert$
boils down to the pure state version of the usual concurrence\cite{CKW,Gittings}.

Now take an element $|\psi\rangle$ of $\mathcal{F}^-$ parametrized as
\beq
|\psi\rangle=v_i (f^i)^+|0\rangle + \frac{1}{3!}P_{ijk}(f^i)^+(f^j)^+(f^k)^+|0\rangle.
\eeq
For this we have
\beq
(\psi,\psi)=\frac{1}{3}v_iP_{jkl}\epsilon^{ijkl},
\eeq
showing us in particular that for a three fermion state with $v_i=0$ no entanglement can occur because of the duality $\wedge^3 \mathbb{C}^4 \cong \wedge^1 \mathbb{C}$. The space $\mathcal{F}^+$ contains two Spin orbits other than the zero vector\cite{Igusa} one with $(\psi,\psi)=0$ and one with $(\psi,\psi)\neq 0$.

\subsection{Five state system}

In the case of an odd dimensional single particle space $\mathcal{F}^+$ is dual to $\mathcal{F}^-$ so one only has to consider one of them. In this case the pairing \eqref{eq:pairing} is only nonvanishing between $\mathcal{F}^+$ and $\mathcal{F}^-$. This allows a slighly different construction than the moment map. We associate an element of $W\oplus W^*$ to an element of $\mathcal{F}^+$ parametrized as
\beq
|\phi\rangle = \eta|0\rangle + \frac{1}{2!}\xi_{ij}(f^i)^+(f^j)^+|0\rangle + \frac{1}{4!} \chi^n \epsilon_{nijkl}(f^i)^+(f^j)^+(f^k)^+(f^l)^+|0\rangle,
\eeq
in the following way.
\beq
\begin{aligned}
 |\phi\rangle \in \mathcal{F}^+ &\mapsto v_{\phi}\in W\oplus W^* \\
 (u,v_{\phi})&=(u \phi,\phi), \; \; \forall u\in W\oplus W^*,
\end{aligned}
\eeq
where the product on the left is the one defined in \eqref{eq:Winnerprod} while the one on the right is the one defined in \eqref{eq:pairing}. A short calculation shows that the components of $v_\phi$ are
\beq
\begin{aligned}
(v_\phi)_i=(f_i \phi,\phi)=2\xi_{ik}\chi^k, && (v_\phi)^i=((f^i)^+ \phi,\phi)=\frac{1}{12}\eta \chi^i-\frac{1}{4}\xi_{jk}\xi_{lm}\epsilon^{ijklm},
\end{aligned}
\eeq
and by construction it transforms as an $SO(10,\mathbb{C})$ vector under $Spin(10,\mathbb{C})$. A quartic invariant can be constructed as $(v_\phi,v_\phi)$ but this turns out to be identically zero. However, $\mathcal{F}^+$ consists of two orbits\cite{Igusa} one with $v_\phi=0$ and one with $v_\phi\neq 0$. 

\subsection{Six state system}

Here $\mathcal{H}=\mathbb{C}^6$ and the Fock space is of dimension 64. We begin with the 32 dimensional even particle subspace $\mathcal{F}^+$. 

\subsubsection{Even particle subspace}

We parametrize a general state with two complex scalars $\eta$, $\xi$ and two antisymmetric $6\times 6$ complex matrices $y$ and $x$ in the following way
\beq
\begin{aligned}
\label{eq:evenparticlestate}
|\phi\rangle =&\eta |0\rangle + \frac{1}{2!} y_{ab} (f^a)^+(f^b)^+|0\rangle + \frac{1}{2!4!}x^{ab}\epsilon_{abcdef} (f^c)^+(f^d)^+(f^e)^+(f^f)^+|0\rangle \\ &+ \frac{1}{6!}\xi \epsilon_{abcdef}(f^a)^+(f^b)^+(f^c)^+(f^d)^+(f^e)^+(f^f)^+|0\rangle.
\end{aligned}
\eeq
Using \eqref{eq:momentexpl} we can calculate the elements of the moment map
\beq
\begin{aligned}
{[A_\phi]}^i_{\;k} &= 2 x^{ia}y_{ak}-\left(\frac{1}{2}\text{Tr}(xy)+\eta\xi\right)\delta^i_{\;k},\\
 [B_\phi]_{jk} &=\frac{1}{4}x^{ab}x^{cd}\epsilon_{abcdjk}-2\xi y_{jk}, \\
 [\beta_\phi]^{il} &= \frac{1}{4}y_{ab}y_{cd}\epsilon^{abcdil}-2\eta x^{il}.
\end{aligned}
\eeq
An easy calculation shows that we have a non-trivial quartic invariant
\beq
\label{eq:quartic1}
\frac{1}{6}q_2(\eta,y,x,\xi)= \left(\eta\xi+\frac{1}{2}\text{Tr}(yx)\right)^2+4\eta \text{Pf}(x)+4\xi \text{Pf}(y)-\frac{1}{2}\left((\text{Tr}(yx))^2-2\text{Tr}(yxyx)\right),
\eeq
where we have introduced the Pfaffian of an antisymmetric matrix $\text{Pf}(x)=\frac{1}{3!2^3}\epsilon_{abcdef}x^{ab}x^{cd}x^{ef}$. We have shown in a previous paper\cite{levsar} that the action of $Spin(12,\mathbb{C})$ on $\mathcal{F}^+$ can be described in the language of the so called Freudenthal triple systems\cite{Freudenthal,Krutelevich} widely known in mathematical and supergravity literature. Particulary $\mathcal{F}^+$ as a vector space is isomorphic to the Freudenthal triple system over the biquaternions. An element of this is parametrized by two complex scalars and two biquaternion entry quaternion-Hermitian $3\times 3$ matrices. The action of $Spin(12,\mathbb{C})$ on $\mathcal{F}^+$ is then just the action of the conformal group of the cubic Jordan algebra $\text{Herm}(3,\mathbb{H})\otimes \mathbb{C}$ on the Freudenthal system. Any Freudenthal triple system admits a quartic invariant and an antisymmetric bilinear product. The quartic invariant is just $\frac{1}{6}q_2(\phi)$ the bilinear product 
is just the 
pairing $(\phi,\psi)$. Every 
element of a Freudenthal triple system has a so called Freudenthal dual which is cubic in the original parameters. This dual is just mapped to $T_\phi |\phi\rangle$. And finally a Freudenthal system allways has five orbits under the action of its conformal group\cite{Krutelevich} thus we can deduce that $\mathcal{F}^+$ is split into five orbits under the action of $Spin(12,\mathbb{C})$. These are:
\begin{enumerate}
 \item rank $|\phi\rangle= 4$ if $q_2(\phi)\neq 0$,
 \item rank $|\phi\rangle= 3$ if $q_2(\phi)= 0$ but $T_\phi |\phi\rangle \neq 0$,
 \item rank $|\phi\rangle= 2$ if $T_\phi |\phi\rangle = 0$ but $T_\phi\neq 0$,
 \item rank $|\phi\rangle= 1$ if $T_\phi  = 0$ but $|\phi\rangle \neq 0$,
 \item rank $|\phi\rangle= 0$ if $|\phi\rangle=0$.
\end{enumerate}
The canonical form of an element of the GHZ-like first orbit is
\beq
|\phi_0\rangle=|0\rangle + (f^1)^+(f^2)^+(f^3)^+(f^4)^+(f^5)^+(f^6)^+|0\rangle,
\eeq
i.e. $\eta=\xi=1$ and $x=y=0$. For this we have $\frac{1}{6}q_2(\phi_0)=1$. The fourth class is the one of pure spinors. One can easily check that for example a state of the form $e^{-B}|0\rangle$ has vanishing moment map.

We can list representatives from all of the classes. Consider a state parametrized by four complex numbers $a,b,c,d$ defined by
\beq
\label{eq:evencanonical}
\begin{aligned}
\eta=0, && y=\left(
\begin{array}{cccccc}
0 & a & 0 & 0 & 0 & 0 \\
-a & 0 & 0 & 0 & 0 & 0 \\
0 & 0 & 0 & b & 0 & 0\\
0 & 0 & -b & 0 & 0 & 0\\
0 & 0 & 0 & 0 & 0 & c\\
0 & 0 & 0 & 0 & -c & 0\\
\end{array}
\right),&& x=0, && \xi=d. 
\end{aligned}
\eeq
For these states we have $\frac{1}{6}q_2(\phi_0)=4\xi\text{Pf}(x)=4 abcd$. The values of the four parameters for the different classes can be found in TABLE \ref{tab:1}.

\begin{table}[h!]
\centering
\begin{tabular}{|c|c|cccc|}
\hline
rank $|\phi\rangle$ & rank $\mathcal{T}_\phi$ & $a$ & $b$ & $c$ & $d$ \\ \hline \hline
4 & 12 & 1 & 1& 1 & 1\\
3 & 6 & 1 & 1& 1 & 0 \\
2 & 2 & 1 & 1& 0 & 0 \\
1 & 0 & 1 & 0& 0 & 0 \\
0 & 0 & 0 & 0& 0 & 0 \\
 \hline
\end{tabular}
\caption{Canonical forms of the orbits of $\mathcal{F}^+$ in six dimensions. The parametrization is given in \eqref{eq:evencanonical}.}
\label{tab:1}
\end{table}

\subsubsection{Odd particle subspace}

Now consider $\mathcal{F}^-$. A general state can be parametrized as
\beq
|\psi\rangle=u_a (f^a)^+|0\rangle + \frac{1}{3!}P_{abc}(f^a)^+(f^b)^+(f^c)^+|0\rangle + \frac{1}{5!} w^l \epsilon_{labcde}(f^a)^+(f^b)^+(f^c)^+(f^d)^+(f^e)^+|0\rangle,
\eeq
where $u$ and $w$ are six dimensional complex vectors and $P$ is a rank 3 antisymmetric tensor. Using \eqref{eq:momentexpl} the moment map reads as
\beq
\begin{aligned}
{[A_\phi]}^i_{\;k} &= 2w^i u_k -(K_P)^i_{\;k}-w^a u_a \delta^i_{\;k},\\
 [B_\phi]_{jk} &=2P_{akj}w^a, \\
 [\beta_\phi]^{il} &= \frac{2}{3!}u_aP_{bcd}\epsilon^{ilbcda},
\end{aligned}
\eeq
where we have defined the matrix
\beq
(K_P)^i_{\;k}=\frac{1}{3!2!}P_{kab}P_{cde}\epsilon^{iabcde},
\eeq
which is important in the ordinary SLOCC classification of the three particle subspace. The non-trivial quartic invariant is
\beq
q_2(\psi)=6(w^iu_i)^2-4 w^i u_j (K_P)^j_{\;i} + \text{Tr} K_P^2.
\eeq
We note that for simple three fermion states of the form
\beq
\label{eq:threefermion}
|\psi_0\rangle =\frac{1}{3!}P_{abc}(f^a)^+(f^b)^+(f^c)^+|0\rangle,
\eeq
we have $q_2(\psi_0)=\text{Tr} K_P^2$ which is just the usual quartic invariant of three fermions with six single particle states\cite{levvran1}. Moreover rank $\mathcal{T}_{\psi_0}=2\cdot \text{rank} K_P$ and $\text{rank}K_P$ is enought to resolve all the SLOCC classes, namely if $\text{rank}K_P=6,3,1,$ or $0$ then $|\psi_0\rangle$ belongs to the GHZ, W, biseparable or separable class respectively\cite{levvran1,levsar2}. 

\subsubsection{Embedded three qubit system}

Recall that an unnormalized three qubit pure state is an element of $\mathbb{C}^
2 \otimes \mathbb{C}^2 \otimes \mathbb{C}^2$ and can be written with the help of
eight complex amplitudes $\Phi_{ijk}$ as
\beq
|\Phi\rangle = \sum_{i,j,k\in \lbrace 0,1\rbrace} \Phi_{ijk} |i\rangle \otimes |
j\rangle \otimes |k\rangle,
\label{3qbitstate}
\eeq
where $|0\rangle,|1\rangle$ are the basis vectors of $\mathbb{C}^2$. The SLOCC 
group of the distinguishable three qubit system\cite{Bennett,Dur} is 
$GL(2,\mathbb{C})
\otimes GL(2,\mathbb{C}) \otimes GL(2,\mathbb{C})$ which acts 
on the amplitudes
as
 \beq
 \begin{aligned}
 \Phi_{ijk}&\mapsto (G_1)_i^{\;\;i'}(G_2)_j^{\;\;j'}(G_3)_k^{\;\;k'}\Phi_{i'j'k'}
 , \\ G_1\otimes G_2\otimes G_3 &\in GL(2,\mathbb{C})\otimes GL(2,\mathbb{C}) 
\otimes
GL(2,\mathbb{C}).
\end{aligned}
\eeq
In this section we show that this system can be embedded in both the even and odd particle subspaces of the six single particle state Fock space with the SLOCC group being a subgroup of the ${\rm Spin}(12,{\mathbb C})$ group. However, unlike in the odd particle subspaces
where the SLOCC group shows up as the expected subgroup of $GL(6,{\mathbb C})\subset {\rm Spin}(12,{\mathbb C})$, in the even particle subspace this group is arising also from taking into account the Bogoliubov transformations {\it not} belonging to the trivial $GL(6,{\mathbb C})$ subgroup.

Consider first the odd particle subspace. If we restrict ourselves to states of the form \eqref{eq:threefermion} and consider only the particle number conserving subgroup $GL(6,\mathbb{C})\subset Spin(12,\mathbb{C})$ we will end up with the usual SLOCC classification of three fermions with six single particle states. Now it is well known that three qubits can be embedded in this system\cite{levvran1,levvran2,djok1,djok2} as
\beq
|P_\Phi\rangle =
\sum_{i,j,k\in \lbrace 0,1\rbrace}
\Phi_{ijk} (f^{i+1})^+ (f^{j+3})^+ (f^{k+5})^+|0\rangle \in \wedge^3 \mathcal{H}.
\label{3qubitodd}
\eeq
Now the three qubit SLOCC group as a subgroup of $GL(6,\mathbb{C})$ is parametrized as

\beq
G=
\left(
\begin{array}{ccc}
G_1 & & \\
& G_2& \\
& & G_3
\end{array}
\right)
\in GL(6,\mathbb{C}).
\eeq
This way the five entanglement classes of three qubits\cite{Dur} are mapped exactly into the five entanglement classes of three fermions\cite{levvran1,levvran2,djok1,djok2}.

Now we will also consider the even particle subspace. 
Our aim is to present the "bosonic" (even number of particles) analogue $\vert\phi_{\Phi}\rangle$ of the "fermionic" state (odd number of particles) $\vert P_{\Phi}\rangle$. Taken together with our previous case in this way we will be given a {\it  dual} description of our three-qubit state $\vert \Phi \rangle$. As the upshot of these considerations we will see that the usual SLOCC group on three-qubits can be recovered from two wildly different
physical situations. 

To this end in view we first calculate the general form of the spin transformations generated by \eqref{eq:T} with only $B$, $\beta$ or $A$ not being zero on a state parametrized as in Eq. \eqref{eq:evenparticlestate}. For $e^{-B}$ we have
\beq
\label{eq:Btrans}
e^{-B}: 
\left(
\begin{array}{c}
\eta \\
y_{ab} \\
x^{cd} \\
\xi
\end{array}
\right)
\mapsto
\left(
\begin{array}{c}
\eta \\
y_{ab}-\eta B_{ab} \\
x^{cd} + \frac{1}{2}\eta  (B\times B)^{cd} -(B\times y)^{cd} \\
\xi-\eta \text{Pf}(B)-\frac{1}{4} \text{Tr}((B\times B)y)+\frac{1}{2}\text{Tr}(Bx)
\end{array}
\right),
\eeq
and for $e^{-\beta}$ we have
\beq
\label{eq:betatrans}
e^{-\beta}: 
\left(
\begin{array}{c}
\eta \\
y_{ab} \\
x^{cd} \\
\xi
\end{array}
\right)
\mapsto
\left(
\begin{array}{c}
\eta+\xi \text{Pf}(\beta)-\frac{1}{4} \text{Tr}((\beta\times \beta)x)-\frac{1}{2}\text{Tr}(\beta y) \\
y_{ab} + \frac{1}{2}\xi (\beta \times \beta)_{ab} +(\beta \times x)_{ab} \\
x^{cd}+\xi \beta^{cd} \\
\xi
\end{array}
\right).
\eeq
Here we have introduced the notation $(C\times D)^{ab}=\frac{1}{4}\epsilon^{abcdef}C_{cd}D_{ef}$. Finally, using \eqref{eq:slocc} and the formula $\epsilon_{abcdef}G^a_{\;a'}...G^f_{\;f'}=(\det G) \epsilon_{a'b'c'd'e'f'}$ we see that the transformation $e^{A-\frac{1}{2}\text{Tr}A}$ acts as
\beq
\label{eq:Atrans}
e^{A-\frac{1}{2}\text{Tr}A}: 
\left(
\begin{array}{c}
\eta \\
y_{ab} \\
x^{cd} \\
\xi
\end{array}
\right)
\mapsto
\left(
\begin{array}{c}
(\det G)^{-\frac{1}{2}}\eta \\
(\det G)^{-\frac{1}{2}}G^{a'}_{\;a}G^{b'}_{\;b}y_{a'b'} \\
(\det G)^{\frac{1}{2}}(G^{-1})^{c}_{\;c'}(G^{-1})^{d}_{\;d'}x^{cd} \\
(\det G)^{\frac{1}{2}}\xi
\end{array}
\right),
\eeq
where the $6\times 6$ matrix $G^a_{\;b}$ is the matrix exponential of the coefficient matrix $A^a_{\;b}$.

Let us now define the state $|\phi_\Phi\rangle$. For this we plug in Eq.(\ref{eq:evenparticlestate}) the eight amplitudes of the three qubit state $|\Phi\rangle$  of Eq. (\ref{3qbitstate}) as 
\beq
\label{eq:embedded3qubitseven}
\begin{aligned}
\eta =\Phi_{000}, && y =\left(
\begin{array}{cccccc}
0 & 0 & 0 & \Phi_{100} & 0 & 0 \\
0 & 0 & 0 & 0 & \Phi_{010} & 0 \\
0 & 0 & 0 & 0 & 0 & \Phi_{001}\\
-\Phi_{100} & 0 & 0 & 0 & 0 & 0\\
0 & -\Phi_{010} & 0 & 0 & 0 & 0\\
0 & 0 & -\Phi_{001} & 0 & 0 & 0\\
\end{array}
\right),\\ 
\xi =-\Phi_{111}, &&
x =
\left(
\begin{array}{cccccc}
0 & 0 & 0 & -\Phi_{011} & 0 & 0 \\
0 & 0 & 0 & 0 & -\Phi_{101} & 0 \\
0 & 0 & 0 & 0 & 0 & -\Phi_{110}\\
\Phi_{011} & 0 & 0 & 0 & 0 & 0\\
0 & \Phi_{101} & 0 & 0 & 0 & 0\\
0 & 0 & \Phi_{110} & 0 & 0 & 0\\
\end{array}
\right)
.
\end{aligned}
\eeq
A very important feature of this embedding is that the mapping between the five  well-known three qubit SLOCC classes, namely the GHZ, W, Bisep, Sep and Null\cite{Dur} and the Spin group orbits of TABLE \ref{tab:1}. is one to one.
Now with the use of equations \eqref{eq:Btrans}, \eqref{eq:betatrans} and \eqref{eq:Atrans} one can easily check that the $B$-transformation generated by
\beq
B =
\left(
\begin{array}{cccccc}
0 & 0 & 0 & -a & 0 & 0 \\
0 & 0 & 0 & 0 & -b & 0 \\
0 & 0 & 0 & 0 & 0 & -c\\
a & 0 & 0 & 0 & 0 & 0\\
0 & b & 0 & 0 & 0 & 0\\
0 & 0 & c & 0 & 0 & 0\\
\end{array}
\right),
\eeq
simply implements the three qubit SLOCC transformation
\beq
G_B=
\left(
\begin{array}{cc}
1 & a\\
0 & 1
\end{array}
\right)
\otimes
\left(
\begin{array}{cc}
1 & b\\
0 & 1
\end{array}
\right)
\otimes 
\left(
\begin{array}{cc}
1 & c\\
0 & 1
\end{array}
\right),
\eeq
while the $\beta$-tranformation generated by
\beq
\beta =
\left(
\begin{array}{cccccc}
0 & 0 & 0 & a & 0 & 0 \\
0 & 0 & 0 & 0 & b & 0 \\
0 & 0 & 0 & 0 & 0 & c\\
-a & 0 & 0 & 0 & 0 & 0\\
0 & -b & 0 & 0 & 0 & 0\\
0 & 0 & -c & 0 & 0 & 0\\
\end{array}
\right),
\eeq
implements
\beq
G_\beta=
\left(
\begin{array}{cc}
1 & 0\\
a & 1
\end{array}
\right)
\otimes
\left(
\begin{array}{cc}
1 & 0\\
b & 1
\end{array}
\right)
\otimes 
\left(
\begin{array}{cc}
1 & 0\\
c & 1
\end{array}
\right).
\eeq
Finally the $A$-transformation generated by
\beq
A =
\left(
\begin{array}{cccccc}
\log a^{-1} & 0 & 0 & 0 & 0 & 0 \\
0 & \log b^{-1} & 0 & 0 & 0 & 0 \\
0 & 0 & \log c^{-1} & 0 & 0 & 0\\
0 & 0 & 0 & \log a^{-1} & 0 & 0\\
0 & 0 & 0 & 0 & \log b^{-1} & 0\\
0 & 0 & 0 & 0 & 0 & \log c^{-1}\\
\end{array}
\right),
\eeq
implements the SLOCC transformations
\beq
G_A=
\left(
\begin{array}{cc}
a & 0\\
0 & a^{-1}
\end{array}
\right)
\otimes
\left(
\begin{array}{cc}
b & 0\\
0 & b^{-1}
\end{array}
\right)
\otimes 
\left(
\begin{array}{cc}
c & 0\\
0 & c^{-1}
\end{array}
\right).
\eeq
This way with succesive $B$, $\beta$ and $A$ transformations we can generate the group $SL(2,\mathbb{C})\otimes SL(2,\mathbb{C}) \otimes SL(2,\mathbb{C})$ acting properly on three qubit amplitudes\cite{Dur}. 
Note however, that in order to reproduce the full SLOCC group $GL(2,\mathbb{C})\otimes GL(2,\mathbb{C}) \otimes GL(2,\mathbb{C})$ in this dual picture we need to extend the Spin group
to ${\mathbb C}^{\times}\times {\rm Spin}(12,{\mathbb C})$.

It is also important to realize that when evaluated at $|\phi_\Phi\rangle$, the quartic invariant of \eqref{eq:quartic1} is just Cayley's hyperdeterminant showing up in the definition of the three-tangle well-known from studies concerning three-qubit entanglement\cite{CKW}. Notice also that although the $B$ and $\beta$ transforms are of Bogoliubov type and hence they change the particle number in the original picture however, now in spite of this they still implement {\it ordinary SLOCC transformations}. 
For these Bogoliubov type transformations
this observation gives rise to a conventional entanglement based interpretation.

\begin{figure}[htbp]
\begin{center}

\includegraphics[width=0.5\textwidth]{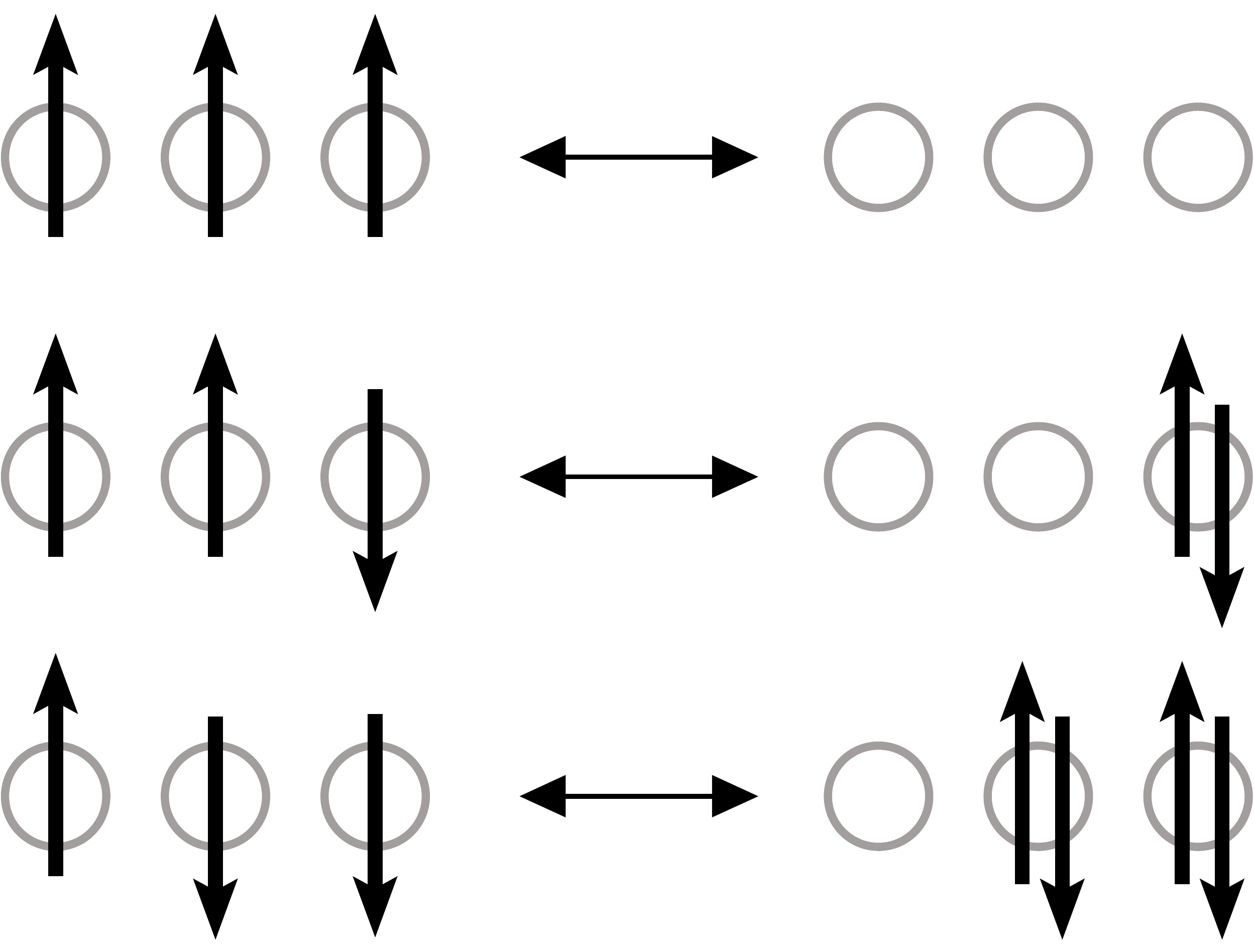}

\caption{Three qubit states $|000\rangle$, $|001\rangle$ and $|011\rangle$ as single occupancy states in a three fermion three node system (left) and as double occupancy states in an even fermion number three node system (right).}
\label{figure:3qubit}
\end{center}
\end{figure}

Let us illustrate this interesting duality between the two different representations $\vert P_{\Phi}\rangle$ and $\vert\phi_{\Phi}\rangle$ of our three-qubit state $\vert \Phi\rangle$ by a set-up borrowed from solid state physics\cite{Eckert,Djokovics}. Suppose we have three nodes where spin $1/2$ fermions can be localized. The states associated to these nodes will be denoted by $|r_1\rangle$, $|r_2\rangle$, $|r_3\rangle$. They are representing a fermion localized on the first, second or the third node. These span a three dimensional complex Hilbert space denoted by $\mathcal{H}_{\text{node}}$. The two-dimensional Hilbert space of a spin $1/2$ is spanned by $\vert\uparrow\rangle$, $\vert\downarrow\rangle$ and is denoted by $\mathcal{H}_{\text{spin}}$. Thus the single particle Hilbert space is the six dimensional one: $\mathcal{H}=\mathcal{H}_{\text{node}}\otimes \mathcal{H}_{\text{spin}}$. 

Moving to multi particle states the creation operator 
associated to the basis 
$|r_i\rangle \otimes |\sigma\rangle$ is $(f^{r_i,\sigma})^+$.
Let us introduce a short-hand notation. 
We will use simple numbers to denote the nodes with spin up and numbers with bars to denote nodes with spin down, i.e. $(f^{r_i,\uparrow})^+=(f^{i})^+$ and $(f^{r_i,\downarrow})^+=(f^{\bar i})^+$, $i=1,2,3$. The connection with the labeling used in \eqref{eq:evenparticlestate} is simply $1\leftrightarrow 1$, $2\leftrightarrow 2$, $3\leftrightarrow 3$, $4\leftrightarrow \bar 1$, $5\leftrightarrow \bar 2$, $6\leftrightarrow \bar 3$. With this notation when embedding three qubits in a three fermion system as in Eq.(\ref{3qubitodd})  after a relabelling of indices as $(i+1,j+3,k+5)\mapsto (3i+1,3j+2,3k+3)$ one can set up the following correspondence between basis states: 
\beq
\begin{aligned}
|000\rangle \leftrightarrow (f^1)^+ (f^2)^+(f^3)^+|0\rangle, && |001\rangle \leftrightarrow (f^1)^+ (f^2)^+(f^{\bar 3})^+|0\rangle, \\
|011\rangle \leftrightarrow (f^1)^+ (f^{\bar 2})^+(f^{\bar 3})^+|0\rangle, 
&& \text{etc.}
\end{aligned}
\eeq
This clearly shows that the three qubit states are mapped to the \textit{single occupancy} states of the three fermion system\cite{Djokovics}. In this case on every site we can find at most one fermion (see: left side of FIG \ref{figure:3qubit}.). On the other hand when the embedding is made into the even particle Fock space as in \eqref{eq:embedded3qubitseven} the correspondence between states reads as
\beq
\begin{aligned}
 |000\rangle &\leftrightarrow |0\rangle, && |001\rangle \leftrightarrow [(f^3)^+ (f^{\bar 3})^+]|0\rangle, \\
|011\rangle &\leftrightarrow [(f^2)^+ (f^{\bar 2})^+][(f^3)^+ (f^{\bar 3})^+]|0\rangle, 
&& \text{etc.}
\end{aligned}
\eeq
This shows that the three qubit states are mapped to the \textit{double occupancy} states of the even particle number Fock space. In this case only empty and twice occupied nodes are allowed (see: right side of FIG \ref{figure:3qubit}.).

\subsubsection{Relation to wrapped membrane configurations in string theory}

Let us also mention that the duality as described above has a particularly nice realization in string theory. Indeed, after invoking some ideas of the recently
discovered Black-Hole/Qubit Correspondence\cite{review} one can also map the two possible embeddings $\vert P_{\Phi}\rangle$ and $\vert\phi_{\Phi}\rangle$ of a three-qubit state $\vert\Phi\rangle$ to the so called IIA and IIB duality frames of toroidal compactifications of type IIA and IIB string theory.
These theories are consistent merely in ten dimensions\cite{Becker}, hence to account for the four space-time dimensions we see six dimensions have to be compactified to tiny six dimensional tori $T^6$. One can regard this six-torus as $T^6=T^2\times T^2\times T^2$ hence the three sites
of Fig. 1. in this picture correspond to three two-tori.
Now recall that these theories are featuring extended objects called (mem)branes that can be interpreted as qubits\cite{wrapornottowrap}. In the case of IIB string theory the situation is depicted by the left hand side of Fig. 1. In this case the theory contains three-branes which correspond to our three-fermions. These branes are wrapping around the noncontractible cycles of $T^2\times T^2\times T^2$. A single $T^2$  contains two basic noncontractible loops.
These loops are corresponding to the two possibilities of spin up and spin down.
Hence the basis states on the left hand side of Fig. 1. give a nice mnemonic for
the basic wrapping configurations of three-branes in the IIB picture.
The three-qubit amplitudes multiplying these basis states are the integers corresponding to the winding numbers.
In the usual four space-time dimensional physics they are reinterpreted
as {\it charges} of electric and magnetic type.
If we also allow the {\it shapes} of these tori to fluctuate in such a way that the volumes are unchanged the natural basis states will be modified. In this case the amplitudes of the three-qubit states turn out to be complex 
depending on the deformation parameters and the winding numbers\cite{qubitsfromextra}.
Hence one can summarize this interpretation of the left hand side of Fig. 1. as: sites with possible spin projections correspond to two-tori taken together with their basic loops, and the three fermions correspond to three-branes. 

In the case of IIA string theory we have $0$-branes, $2$-branes, $4$-branes and $6$-branes. Now they correspond to states containing $0,2,4,6$ fermions in a Fock space. 
Clearly unlike in the previous case now the number of fermions is {\it not} fixed.
The membranes of different dimensions can wrap different dimensional subspaces of $T^2\times T^2\times T^2$. Clearly the vacuum state corresponds to a $0$-brane not wrapping any volume. $2$-branes are wrapping any of the volumes of the three $2$-tori $T^2$ etc. Generally we have $2n$-branes wrapping the $2n$-volumes of $T^{2n}$ tori with $n=0,1,2,3$. 
Some of the basis configurations are illustrated by the states on the right hand side of Fig.1. 
Again the amplitudes multiplying these basis states are integers corresponding to wrapping numbers facilitating an interpretation as electric and magnetic charges in the usual four-dimensional space-time picture.
Note that in this case the {\it volumes} of the sites i.e. the two-tori are subject to fluctuations\cite{Becker}.
Again in a convenient basis adapted to volume fluctuations we will be given three-qubit states with complex amplitudes instead depending on the deformation parameters and wrapping numbers.

The two different embeddings of the {\it same} SLOCC subgroup (the so clalled U-duality group)
$SL(2,{\mathbb Z})\times SL(2,{\mathbb Z})\times SL(2,{\mathbb Z})$
give a nice example of the two wildly different physical scenarios that are amenable to a three-qubit interpretation.
It is also important to realize that under duality the deformation parameters of {\it shape} and {\it volume} of the tori are also exchanged.
This is an example of the famous {\it mirror symmetry}\cite{Becker} well-known in string theory. 
Is there an analogue of this phenomenon in solid state physics?
For an implementation of this idea clearly the amplitudes of the states
$\vert P_{\Phi}\rangle$ and $\vert\phi_{\Phi}\rangle$
should depend somehow on additional parameters with peculiar properties specified by some extra physical constraints.

\section{Conclusions}
\label{sec:conclusions}

In this paper we proposed a generalization of the usual SLOCC and LU classification of entangled quantum systems represented by fermionic Fock spaces. Our approach is based on the representation theory of the Spin group.  In particular for classifying the entanglement types of fermionic systems with $d$ single particle states and an indefinite number of particles
we suggested the group $Spin(2d,\mathbb{C})$.
Our new classification scheme naturally incorporates the usual one based on fixed particle number subspaces. However, our approach also gives rise to the notion of equivalence under transformations that are changing
the particle number.
As far as entanglement is concerned, the new ingredient in our scheme is the identification of states that
can be obtained from each other by Bogoliubov transformations.
There are always at least two orbits corresponding to the "fermionic" (odd number of particles) and "bosonic" (even number of particles) subspaces of the full Fock space. Thus entanglement is prohibited between states belonging to these orbits.

The totally unentangled states are represented by the pure spinors. They are incorporating all the possible vacua and the states that can be written in terms of a single Slater determinant. Local processes that conserve probabilities are described by the (fermionic) LU group $U(d)$. The generalization of this 
group should be the 
compact real subgroup $SO(2d,\mathbb{R})$. This group also has the properties mentioned previously but in addition it only incorporates the 
LOCC group and it conserves probabilities. 

In order to enable an explicit construction in the second half of this paper we have presented some useful mathematical tools. Namely, we have introduced notions such as the Mukai pairing and the moment map. 
We have constructed the basic invariants of the Spin group and  have shown how some of the known SLOCC invariants are just their special cases.
As an example we have presented the classification of fermionic systems based on the group ${\rm Spin}(12,{\mathbb C})$ where the underlying single particle Hilbert space has dimension six.
An intriguing duality between the two different possibilities for embedding three-qubit systems inside this fermionic system has been revealed.
We have elucidated this duality via establishing an interesting connection between such embedded three-qubit systems and configurations of wrapped membranes
reinterpreted as qubits.
Finally we mention that for a treatment of the bosonic case it is clear that the group $SO(2d,\mathbb{C})$ has to be replaced by $Sp(2d,\mathbb{C})$ but the Fock space in this case is infinite dimensional which questions the possibility of further development.

Despite how natural this generalization is currently it is not known to us whether there exist {\it realistic} physical scenarios where our newly proposed classification scheme turns out to be useful. 
Such possibilities certainly deserve attention to be fully explored and will be subject to further investigations.

\section{Acknowledgements}                                                      
One of us (P. L.) would like to acknowledge financial support from the MTA-BME  Kondenz\'alt Anyagok Fizik\'aja Kutat\'ocsoport under grant no: 04119.

\end{document}